\documentclass[usenatbib]{mn2e}
\usepackage{graphicx}
\usepackage{amsmath}

\def\apj{\rm ApJ}
\def\apjl{\rm ApJL}

\def\aj{\rm AJ}
\def\mnras{\rm MNRAS}
\def\nat{\rm Nature}

\def\aap{\rm AAP}
\def\araa{\rm ARA\&A}

\def\gax{\mathrel{\raise.3ex\hbox{$>$}\mkern-14mu\lower0.6ex\hbox{$\sim$}}}
\def\lax{\mathrel{\raise.3ex\hbox{$<$}\mkern-14mu\lower0.6ex\hbox{$\sim$}}}
\def\gtorder{\mathrel{\raise.3ex\hbox{$>$}\mkern-14mu
             \lower0.6ex\hbox{$\sim$}}}
\def\ltorder{\mathrel{\raise.3ex\hbox{$<$}\mkern-14mu
             \lower0.6ex\hbox{$\sim$}}}

\def\ha{\hat{\alpha}''_E}

\begin{document}

\title [Over-constrained Gravitational Lens Models and the Hubble Constant]
   {Over-constrained Gravitational Lens Models and the Hubble Constant}

\author[C.~S. Kochanek]{ 
    C.~S. Kochanek$^{1,2}$ 
    \\
  $^{1}$ Department of Astronomy, The Ohio State University, 140 West 18th Avenue, Columbus OH 43210 \\
  $^{2}$ Center for Cosmology and AstroParticle Physics, The Ohio State University,
    191 W. Woodruff Avenue, Columbus OH 43210 \\
   }

\maketitle

\begin{abstract}
It is well known that measurements of $H_0$ from gravitational lens time delays scale as 
$H_0 \propto 1 - \kappa_E$ where $\kappa_E$ is the mean convergence at the Einstein radius 
$R_E$ but that all available lens data other than the delays
provide no direct constraints on $\kappa_E$.  The properties
of the radial mass distribution constrained by lens data are $R_E$ and the dimensionless 
quantity $\xi = R_E \alpha''(R_E)/(1-\kappa_E)$ where $\alpha''(R_E)$ is 
the second derivative of the deflection profile at $R_E$.  Lens models with too few degrees 
of freedom, like power law models with densities $\rho \propto r^{-n}$, have a 
one-to-one correspondence between $\xi$ and $\kappa_E$ (for a power law model,
$\xi=2(n-2)$ and $\kappa_E = (3-n)/2 = (2-\xi)/4$).  This means that highly constrained
lens models with few parameters quickly lead to very precise but inaccurate estimates of 
$\kappa_E$ and hence $H_0$.  Based on experiments with a broad range of plausible dark matter halo 
models, it is unlikely that any current estimates of $H_0$ from gravitational lens 
time delays are more accurate than $\sim 10\%$, regardless of the
reported precision. 
\end{abstract}


\section{Introduction}

\cite{Refsdal1964} pointed out that the time delays between multiple images in a
gravitational lens could be used to determine the Hubble constant.  There was
a long delay before the discovery of the first lensed quasar (\citealt{Walsh1979})
and then considerable controversy over the measurement of the first time
delay (\cite{Schild1990} versus \cite{Press1992}, resolved in favor of the
former by \cite{Kundic1997}).  The measurement of delays is now routine
(e.g., \citealt{Bonvin2019}, \citealt{Courbin2018}, \citealt{Bonvin2017}, recently)
and the estimates are generally robust (e.g., \citealt{Liao2015}).  The
challenge lies in their cosmological interpretation.  Individual lenses
yield estimates of $H_0$ with reported precisions of 4-10\% (see Table 1)
with higher precisions depending on averaging the estimates from large numbers of
lenses.  The present state of the art comes from the H0LiCOW collaboration,
who report a 2.4\% measurement of $H_0$ using six gravitational lenses
(\citealt{Wong2019}). 

The time delay $\Delta t$ in a lens is roughly proportional to 
$H_0^{-1}  \left(1 - \kappa_E \right)$
where $\kappa_E $ is the mean convergence (dimensionless
surface density)\footnote{To be more precise, is is proportional
to $1-\langle \kappa \rangle$ where $\langle \kappa \rangle$ is
the mean surface density in the annulus bounded by the lensed 
images.} at the Einstein radius $R_E$ (\citealt{Kochanek2002}).  
Unfortunately, no gravitational lens observable other than the
time delay directly constrains 
$\kappa_E$ (see, e.g., \citealt{Gorenstein1988},
\citealt{Kochanek2002}, \citealt{Kochanek2006}, \citealt{Schneider2013},
\citealt{Wertz2018}, \citealt{Sonnenfeld2018}), so some
additional constraint on the mass distribution is required to
determine $H_0$ from a time delay.  It was quickly realized
that stellar dynamical measurements, usually just meaning the central velocity 
dispersion, could provide this constraint (e.g., \citealt{Grogin1996},
\citealt{Romanowsky1999}, \citealt{Treu2002}).

If we explore simple lens models constrained by a stellar
velocity dispersion $\sigma_*$, we find that the fractional 
uncertainty in $H_0$ is roughly equal to the fractional uncertainty 
in $\sigma_*^2$.  Since the reported uncertainties in $\sigma_*$ for
the H0LiCOW lenses range from 6-10\% (see Table 1), 
$H_0$ should only be constrained to 12-20\%.  The 
$H_0$ uncertainties reported by H0LiCOW of only 4-10\% (also Table 1) 
are, however, far smaller even after including all other sources of 
uncertainty in the models (e.g., time delays, the local
environment, etc.). This means that the constraints on $\kappa_E$ and 
thus $H_0$ must be coming from the lensing constraints on the 
mass model rather than the stellar dynamical constraints.
In fact, the uncertainties in $H_0$ are so small compared to those
in $\sigma_*$, that the stellar dynamical 
measurements must be making almost no contribution to the overall
estimate of $H_0$.

As already noted, lensing data cannot determine 
$\kappa_E$ -- it is a fundamental degeneracy in the physics of 
gravitational lensing.  Lens models determine $\kappa_E$
only because the mathematical structure of any density model
implies a relationship between the aspects of the model
constrained by lens data and the surface density at the 
Einstein ring.  If the density model has too few degrees of
freedom compared to the number of lensing constraints, then
one quickly obtains a very precise, but likely inaccurate,
constraint on $\kappa_E$ and $H_0$.  Since the errors are
systematic rather than random, there is also no reason to
believe that they are reduced by averaging over multiple
systems.  

Most of these points have been made previously (e.g., \citealt{Kochanek2006}, 
\citealt{Schneider2013}, \citealt{Xu2016}, \citealt{Unruh2017}, 
\citealt{Sonnenfeld2018}).  Here we make these arguments using
a different set of analytic results and numerical experiments,
which clearly show that the simple density models presently used
for most inferences about $H_0$ from gravitational lens time delays
suffer from these problems and become increasingly unreliable as
the reported precision becomes smaller than $\sim 10\%$.  The
arguments are presented in \S2, and we summarize the results in
\S3.

\begin{figure}
\centering
\includegraphics[width=0.50\textwidth]{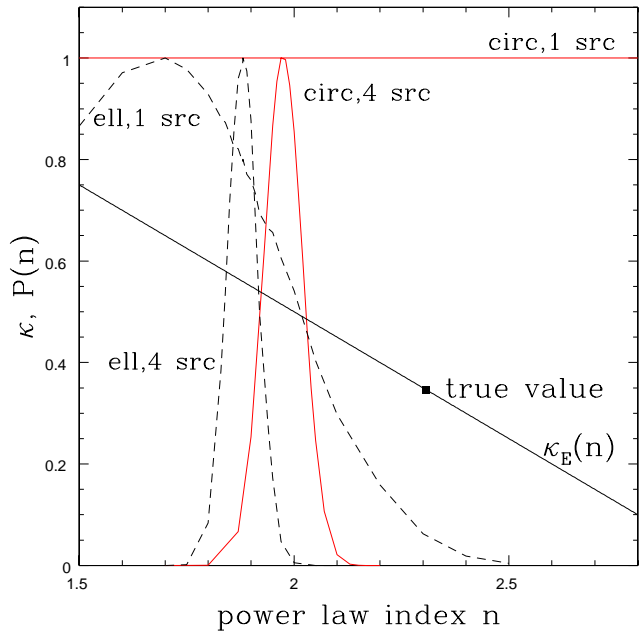}
\caption{Probability of the power law index $n$ for a
 \protect\cite{Hernquist1990} model lens with an Einstein
 radius of $R_E = 1.3s$ for circular (red, solid)
 and ellipsoidal (black, dashed) models fitting 
 either one source producing two images (``1 src'')
 or four sources producing 12 images (``4 src'').
 For the circular lens, matching the values of
 $\xi$ predicts that the best fit power law model 
 should have $n=2$.  The solid line shows the
 dependence of the convergence at the Einstein
 radius $\kappa_E(n)$ on the power law index,
 where the point labeled ``true value'' is the
 correct value for the input
 model. 
  }
\label{fig:modchi}
\end{figure}

\begin{figure}
\centering
\includegraphics[width=0.50\textwidth]{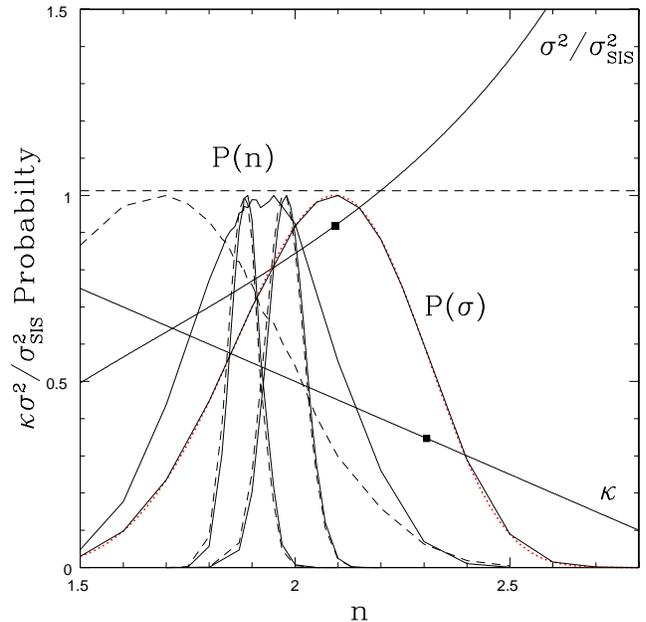}
\caption{ Changes in the probability distributions for $n$
  from Fig.~\protect\ref{fig:modchi} after adding a 10\%
  estimate of the velocity dispersion centered on the
  true value.  The dashed curves are the probability
  distributions from Fig.~\protect\ref{fig:modchi} and 
  the solid curves are the probability distributions 
  including the dynamical constraint.  The $\sigma^2/\sigma_{SIS}^2$
  curve shows the dependence of the velocity dispersion inside
  the aperture $R<s$ compared to the velocity dispersion 
  of an SIS lens model.  
  }
\label{fig:modchi2}
\end{figure}

\begin{table*}
  \centering
  \caption{H0LiCOW Lenses}
  \begin{tabular}{llll}
  \hline
  \multicolumn{1}{c}{Lens}    &
  \multicolumn{1}{c}{$\sigma_*$}   &
  \multicolumn{1}{c}{$H_0$} & 
  \multicolumn{1}{c}{References}\\
  \multicolumn{1}{c}{}    &
  \multicolumn{1}{c}{(km/s)}   &
  \multicolumn{1}{c}{(km/s/Mpc)} \\
\hline
HE0435$-$1223       &$222\pm15$  (7\%) &$73\pm6$  (8\%) &\cite{Wong2017} \\ 
PG~1115$+$080       &$281\pm25$  (9\%) &$83\pm8$ (10\%) &\cite{Tonry1998},\cite{Chen2019}\\ 
RXJ1131$-$1231      &$323\pm20$  (6\%) &$80\pm6$  (8\%) &\cite{Suyu2013} \\ 
SDSS~1206$+$4332    &$290\pm30$ (10\%) &$69\pm4$  (6\%) &\cite{Agnello2016},\cite{Birrer2019} \\ 
B1608$+$656         &$260\pm15$  (6\%) &$71\pm3$  (4\%) &\cite{Suyu2010} \\ 
WFI2033$-$4723      &$250\pm19$  (8\%) &$72\pm4$  (6\%) &\cite{Rusu2019} \\ 
\hline
\end{tabular}
  \label{tab:data}
\end{table*}

\section{The Role of Parameters in the Mass Model}
\label{sec:mass}

H0LiCOW basically uses two mass models for the lenses. The
 first model is a simple power law producing a deflection
angle of $\alpha(\theta) = b^{n-1} \theta^{2-n}$.  The model has two 
parameters, the Einstein radius $b$ and the power-law index $n$
where $n=2$ is the SIS model.   The second model combines the photometric 
model of the lens galaxy with a \cite{Navarro1997} NFW profile
\begin{equation}
   \rho \propto { 1 \over r (a+r)^2 } 
   \label{eqn:nfwprof}
\end{equation}
for the dark matter halo.  In theory, this model has three parameters, a mass
to light ratio for the stellar profile, a density normalization for the
NFW profile and its scale length $a$.  In practice, the scale length is 
constrained by a fairly strong prior to vary by only 10-15\%,  
which effectively makes this a two parameter model as well.

In this section we first review the basic problem that lensing data
mathematically cannot determine the surface density $\kappa_E$
needed to determine $H_0$ and derive the property of lens models
that lens data does constrain.  Next we illustrate the problem
with a specific example of how mass models with small numbers of 
parameters can lead to increasingly precise but inaccurate estimates 
of $\kappa_E$.  Finally, we
show that a range of plausible models for dark matter halos when
modeled using the H0LiCOW mass distributions commonly have fractional systematic
errors in $H_0$ of $10\%$ or more.

\subsection{What Do Lens Models Measure?}

We start by reviewing the basic problem.
Consider a power law model for circular lens with two images at 
$r_1$ and $-r_2$ ($r_1 \leq r_2$).  The lens equations require that
\begin{equation}
  r_1- b^{n-1} r_1^{2-n} = -r_2+b^{n-1} r_2^{2-n}.
\end{equation} 
We can then solve for the Einstein radius,
\begin{equation}
   b^{n-1} = { r_1 + r_2 \over r_1^{2-n} + r_2^{2-n}}.
   \label{eqn:bval}
\end{equation}
Not surprisingly, with only one constraint, a solution can be found
for any power law index.   If we have an additional set of images at
$r_3$ and $-r_4$ ($r_4\leq r_3$ and $r_2 \leq r_4 \leq r_3 \leq r_1$), 
then there is a unique solution from 
solving the transcendental equations
\begin{equation}
 { r_1 + r_2 \over r_1^{2-n} + r_2^{2-n}} =
 { r_3 + r_4 \over r_3^{2-n} + r_4^{2-n}}
\end{equation}
for the power law index.  Because the model has only two parameters,
the mass distribution is now exactly defined everywhere up to the
uncertainties in the position measurements.  In particular, 
the convergence at the Einstein ring is now forced to be
$\kappa_E = (3-n)/2$ which in turn forces a particular value for
$H_0$ given a time delay.  The mass distribution away from the
Einstein ring is also full specified, eliminating any important
constraint from the dynamical data because the fractional uncertainties
in lensing constraints are generally far smaller than the
fractional uncertainties in velocity dispersions.

There are, however, two fundamental problems.
First, as noted in the introduction, whatever the available lensing constraints, 
the one quantity they do not directly constrain is the mean surface
density needed to convert a time delay into $H_0$.  The conversion of the
lensing constraints into a value of $\kappa_E$ is entirely
set by the functional form of the mass model and its flexibility.
Second, the lens geometry has absolutely 
no information on the mass distribution inside or outside
the annulus encompassing the lens images 
-- the exactly determined mass distribution for
these regions is purely an extrapolation set by the functional form
of the mass model. 
  
These two points are also easily demonstrated non-parametrically
(see \citealt{Kochanek2002}, \citealt{Kochanek2006}).  
Let the mass of the lens between two radii be
\begin{equation}
   m(r_1,r_2) = 2 \int_{r_1}^{r_2} u du \kappa(u)
\end{equation}
where $\kappa(r)$ is the convergence (surface density) profile of the lens.  
The deflection angle is then $\alpha(r) =  r^{-1} m(0,r)$ and the lens
equations require that
\begin{equation}
  r_1 - r_1^{-1} m(0,r_1) =
 -r_2 + r_2^{-1} m(0,r_2).
   \label{eqn:constraint}
\end{equation}
Now $m(0,r_1) = m(0,r_2)+m(r_2,r_1)$, so
\begin{equation}
  m(0,r_2) = r_1 r_2 - 
   \langle \kappa\rangle_{21}
   r_2 (r_1-r_2),
   \label{eqn:lens}
\end{equation}
where 
\begin{equation}
   \langle \kappa \rangle_{ij}
  = { 2 \over { r_j^2 - r_i^2 } } \int_{r_i}^{r_j} u du \kappa(u)
\end{equation} 
is the mean convergence in the annulus bounded by $r_1$ and
$r_2$.  For a thin annulus, the Einstein radius is 
$R_E^2= r_1 r_2$ independent of the surface density,
and the mean surface density is the quantity that
determines the $H_0$ given the time delay since
$H_0 \propto 1-\langle \kappa\rangle_{21}$.  Stellar dynamics
essentially provides an independent constraint on $m(0,r_2)$,
thereby allowing an estimate of $\langle \kappa\rangle_{21}$
and hence $H_0$. 

Adding additional lensing constraints does nothing to
remove the degeneracy.  Suppose $r_1$ and $r_2$ bound
the region containing lensed images, and we again add an
additional pair of lensed images with $r_2 < r_4 < r_3 < r_1$.
There is now a second constraint equation like 
Eqn.~\ref{eqn:constraint}.  The non-parametric
parameters of the model are now $m(0,r_2)$, $\langle\kappa\rangle_{24}$,
$\langle\kappa\rangle_{43}$ and $\langle\kappa\rangle_{31}$, leaving
us with four parameters to be constrained by two equations. 
Viewed as a non-parametric model, the number of parameters
expands faster than the number of constraints and the
$H_0$ degeneracy problem cannot be eliminated no matter how
many additional pairs of lensed images are added.  

The annulus encompassing the lensed images of the quasar 
and its host is typically rather narrow, so using a simple
functional form to describe the mass distribution in this annulus
is likely quite reasonable.  The problems are (1) that the constraints 
only apply over the annulus containing the lensed images -- any
prediction of the mass distribution beyond the annulus is 
purely an extrapolation, and (2) that they cannot constrain the
quantity $\kappa_E$ needed to determine $H_0$.  
We can illustrate this by first
determining what property of a lens is constrained by the
data, and then by constructing a model where two radically
different radial mass distributions and predictions for
$H_0$ are essentially indistinguishable using lens data. 

Suppose we locally expand the deflection angle as a Taylor series
near the Einstein radius, $R_E$,
\begin{equation}
   \alpha(r) \simeq R_E + 2(\kappa_E-1)(r-R_E) + {1 \over 2 } \alpha''_E(r-R_E)^2
\end{equation}
where $\kappa_E$ is the convergence and $\alpha''_E$ is the second
derivative of the deflection profile at $R_E$.
The lens equation for a source at radius $\beta$ is then
\begin{equation}
   \beta = -2 (\kappa_E-1)(r-R_E) + { 1\over 2 } \alpha''_E(r-R_E)^2.
\end{equation}
for one image and with the signs flipped on the right side of the
equation for the other image.  We can divide both sides by 
$1-\kappa_E$, to get 
\begin{equation}
   \hat{\beta} = 2 (r-R_E) + { 1\over 2 } \hat{\alpha}''_E(r-R_E)^2.
    \label{eqn:degen}
\end{equation}
where $\hat{\beta}=\beta/(1-\kappa_E)$ and $\hat{\alpha}''_E=\alpha''_E/(1-\kappa_E)$.  
Since the source position
$\beta$ is not an observable, Eqn.~\ref{eqn:degen} means that for 
images near the Einstein ring, lens models determine $\ha$ and
two lens models are indistinguishable if they have the same $\ha$.
Alternatively, we can introduce the dimensionless quantity
\begin{equation}
   \xi = R_E \ha = { R_E \alpha''(R_E) \over 1 - \kappa_E }
\end{equation}
as the second property of the radial mass distribution after $R_E$
that can be well-constrained by lens data.
Because the uncertainties in $R_E$ are generally small, the uncertainties
in $\xi$ will be dominated by the uncertainties in $\ha$.

Many previous studies have found that lens models modeled as
a power law with $\rho \propto r^{-n}$ favor logarithmic slopes
$n \simeq 2$ close to the $n=2$ slope of an isothermal sphere (e.g.,
\citealt{Rusin2005}, \citealt{Gavazzi2007}, \citealt{Koopmans2009},
\citealt{Auger2010}, \citealt{Bolton2012}).
This does not mean that the typical slope of the 
density distribution on the scale of the Einstein radius has 
$n \simeq 2$.  Instead, there is a one-to-one relation that 
$\xi = 2(n-2)$ for the power-law models and the true physical
constraint implied by finding $n \simeq 2$ is that $\xi \simeq 0$.
It is again important to emphasize that lens models do not determine 
$\kappa_E$, the quantity needed to estimate $H_0$. The functional
form chosen for the mass model implies some value of $\kappa_E$ given
the value of $\xi$, but a different mass model will lead to 
a different value of $\kappa_E$ for the same value of $\xi$.
For the power law models, $\kappa_E = (3-n)/2 = (2-\xi)/4$,
with $\kappa_E = 1/2$ for $n=2$ or $\xi=0$.  However, a different
mass model will predict a different value of $\kappa_E$ for the
same value of $\xi$.

\subsection{A Demonstration of the Problem}

Consider the \cite{Hernquist1990} model,
\begin{equation}
  \rho \propto { 1 \over r (s+r)^3 },
\end{equation}
where the scale radius is related to the effective radius by
$s \simeq 0.55 R_e$.  For a \cite{Hernquist1990} model lens,  
the value of $\xi$ depends on the position of the Einstein radius
relative to the break radius $R_E/s$, and $\xi = 0$ for $R_E/s \simeq 1.3$
where $\kappa_E \simeq 0.35$ is the convergence.  If we model this
lens as a power law,  we should find that $n \simeq 2$ with 
$\kappa_E \simeq 0.5$ as the convergence. This means that the
power law lens model will produce a fractional error in 
$H_0$ of $f=H_{true}/H_{model}-1 \simeq 30\%$.    

Figure~\ref{fig:modchi} shows a sequence of four cases fitting this 
example of a \cite{Hernquist1990} lens model with a power law model.  We ignore the 
generation of faint third images by the \cite{Hernquist1990} model
and the flux ratios of the images.  For computing a goodness of
fit, we assume astrometric uncertainties of $0.003s$ for the 
image and lens positions and no constraints on the ellipticity
of the lens or the external shear for the ellipsoidal models.  
The shear and ellipticity
parameters remain reasonable without additional constraints.
We fit the fake data using {\tt lensmodel} (\citealt{Keeton2001}, 
\citealt{Keeton2011}) with the $\chi^2$ goodness of fit computed on the
image plane.

We first considered two circular lens models.
In the first, we place one image at $1.1 R_E$, which
has a second image at $-0.89998 R_E$.  Note that the image
separation of $1.9998 R_E$
is essentially indistinguishable from the $2R_E$
that would be produced by an SIS model.  As seen in 
Fig.~\ref{fig:modchi}, this data can be perfectly fit 
($\chi^2\equiv0$) independent of the slope of the power 
law as expected from Eqn.~\ref{eqn:bval}. For the 
second model, we added three additional sources that
produced outer images at $1.05 R_E$, $1.2R_E$ and 
$1.3 R_E$, respectively.  The separations of the
three resulting image  pairs are also essentially 
indistinguishable from the
$2 R_E$ prediction of an SIS model.  If we fit these
4 image pairs, we now find that the model is strongly
constrained to have $n=2$, as expected from matching the
values of $\xi$. The best model ($n=1.974$) is still a 
perfect fit with $\chi^2=0.015$ for three degrees of freedom.
The surface density at the Einstein ring implied by the
model is, however, completely wrong, leading to a 30\%
error in $H_0$.
Adding more lensing constraints will never solve the problem --
the $\chi^2$ distribution will simply steadily narrow around
$n \simeq 2$ with smaller and smaller uncertainties in both
$n$ and the implied value of $\kappa_E$.  

We next considered the same cases but with an ellipsoidal lens
in an external shear.  We gave the \cite{Hernquist1990} model a 
surface density axis ratio of $q=0.65$ and added an external shear of 
$\gamma=0.05$ at a randomly chosen angle.  For the ellipsoidal
models, we view $s$ as the intermediate axis scale length and
again normalize the mass so that $R_E = 1.3s$ ({\tt lensmodel}
uses the major axis scale length of $s q^{-1/2}$ to define the
models).  We again placed images at $1.05 R_E$, $1.1 R_E$, $1.2 R_E$
and $1.3 R_E$ and random angles around the lens and then found
their companion images. The
two images closer to $R_E$ produced four images, and the two further
from $R_E$ produced two images, so we now have 12 images 
in total. 

We first repeated the fits using the four image system associated
with an image at $1.1 R_E$.  We again find a good fit, but at
$n \simeq 1.7$ with $\chi^2=0.004$. Formally, the model has
fewer constraints than parameters ($-1$ degrees of freedom).
The goodness of fit is not independent of $n$ but clearly selects a
preferred range, albeit with relatively large uncertainties.
We are confident that this is a consequence of the limited 
degrees of freedom in the angular structure of the mass model.  
The density distribution of the \cite{Hernquist1990} model out to
$R_E$ drops more slowly than the $n=2$ power-law model, so
for the same quadrupole it will have larger higher order 
multipoles.  The power-law models compensate by shifting to
lower $n$, less centrally concentrated mass distributions. 
\cite{Kochanek2006} has an extensive discussion on the 
angular structure of lens models.

If we now add in the other three sources and fit all 12 images,
Fig.~\ref{fig:modchi} shows that $n$ is again tightly constrained but still 
offset to lower $n$ than the circular models.  The sense of the
shift only exacerbates the problems for $H_0$, since these models
have surface densities at $R_E$ even higher than the $n=2$ SIS model
and so are still further from the input model.  The best fit models at 
$n\simeq 1.88$ are statistically good fits with $\chi^2=3.3$
for 3 degrees of freedom.
Adding additional sources producing multiple
images simply narrows the probability distribution $P(n)$.

Fig.~\ref{fig:modchi2} shows the consequences of adding dynamical 
constraints to the lensing constraints illustrated in Fig.~\ref{fig:modchi}.
We assume a measured dispersion equal to the true dispersion
for a \cite{Hernquist1990} model inside the aperture $R<s$
with a 10\% uncertainty. 
Fig.~\ref{fig:modchi2} also shows the lensing only probability
distributions for $n$ and the joint lensing$+$dynamics probability
distributions.
The circular model with only one two image lens system plus dynamics comes
closest to yielding models with the correct value of $\kappa_E$
since the joint probability distribution is simply the dynamical
probability distribution as the lens model imposes no constraint
on $n$.  For the elliptical model with one four image lens system,
the dynamical constraint shifts the lensing distribution
to be less inconsistent with the correct value of $\kappa_E$.
For both the circular and ellipsoidal models with 4 sources, the 
probability distributions are essentially unchanged after adding
the dynamical constraint.  The lens model is so strongly constrained
by the lens data, that the relatively weaker dynamical constraints
have little effect.

We tried a broad range of additional numerical examples for a range
of mass models. In circular models, the solution always converges to
match the $\xi$ of the input model.  In ellipsoidal models with 
external shear, there are modest shifts from the $\xi$ of the input
model.  These experiments explain the puzzle discussed in the introduction.
In mass models with few degrees of freedom and
very strong lensing constraints, the lens data ``pins'' the mass
model to match the $\xi$ required by the data.  The weaker dynamical 
constraints then have little effect and estimates of $H_0$
(i.e., $\kappa_E$) show little sensitivity to changes in the
velocity dispersion.  Unfortunately, Figs.~\ref{fig:modchi} and 
\ref{fig:modchi2} also show that the accuracy of the estimate
of $H_0$ was greatly reduced rather than enhanced by the use
of the additional strong lensing constraints.

\begin{figure*}
\centering
\includegraphics[width=0.80\textwidth]{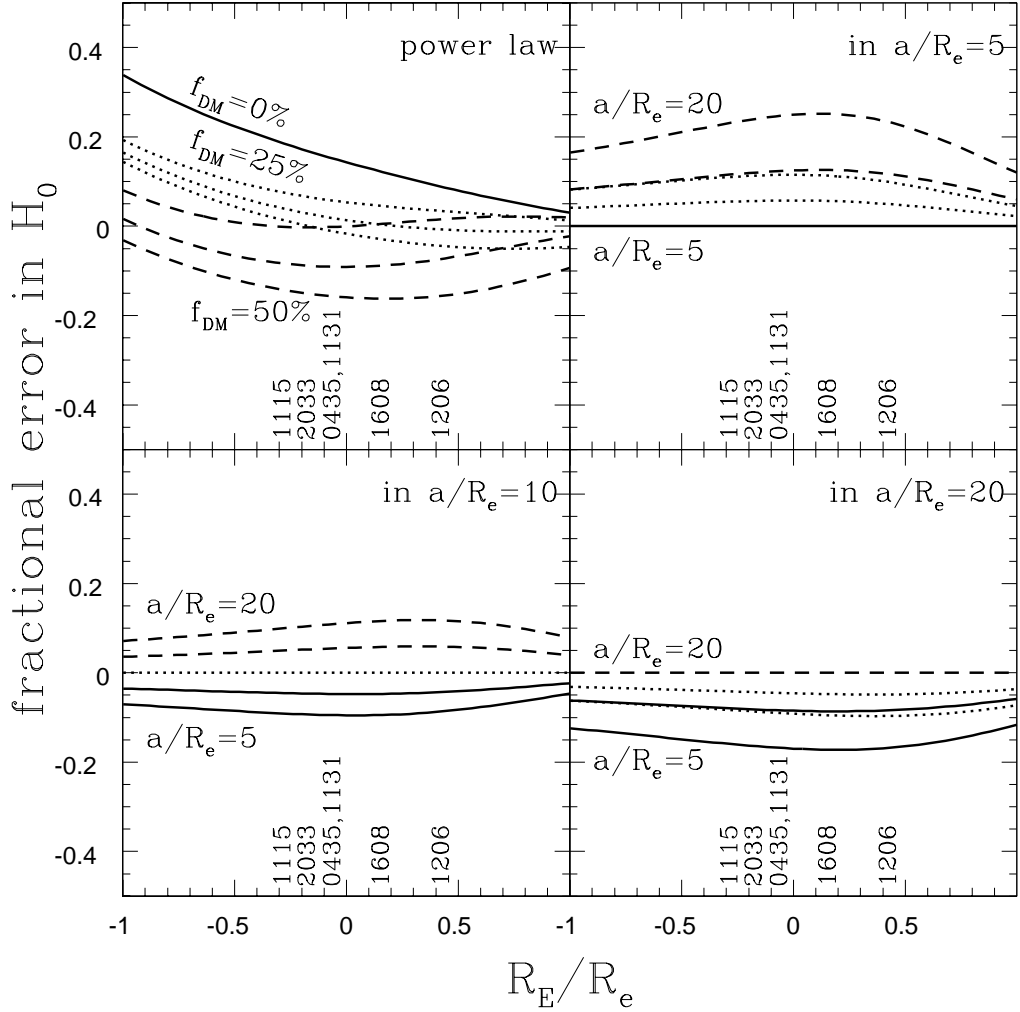}
\caption{ Fractional errors in $H_0$ for an input deV$+$NFW halo
  modeled as a power law (top left) or a deV$+$NFW halo model
  (other panels). In the power law panel the dark matter fraction
  is $f_{DM}=0\%$ (solid), $25\%$ (dotted) or 50\% (dashed) for
  input NFW scale lengths of $a/R_E=5$ (top), $10$ or
  $20$ (bottom).  For the NFW panels, the input NFW scale length is 
  $a/R_E=5$ (top right) $10$ (lower left) or $20$ (lower right), and  
  the model NFW scale length is $a/R_E=5$ (dotted), $10$ (solid)
  or $20$ (dashed) in each panel. Results are shown for dark matter
  fractions of $f_{DM}=25\%$ or $50\%$ with larger fractional errors
  for larger dark matter fractions.  When $f_{DM}=0\%$ or the input
  and model NFW scale lengths are the same, the fractional error is
  zero. The locations of the H0LiCOW lenses in $R_E/R_e$ are indicated by
  the lens names.
  }
\label{fig:nfw}
\end{figure*}

\begin{figure*}
\centering
\includegraphics[width=0.80\textwidth]{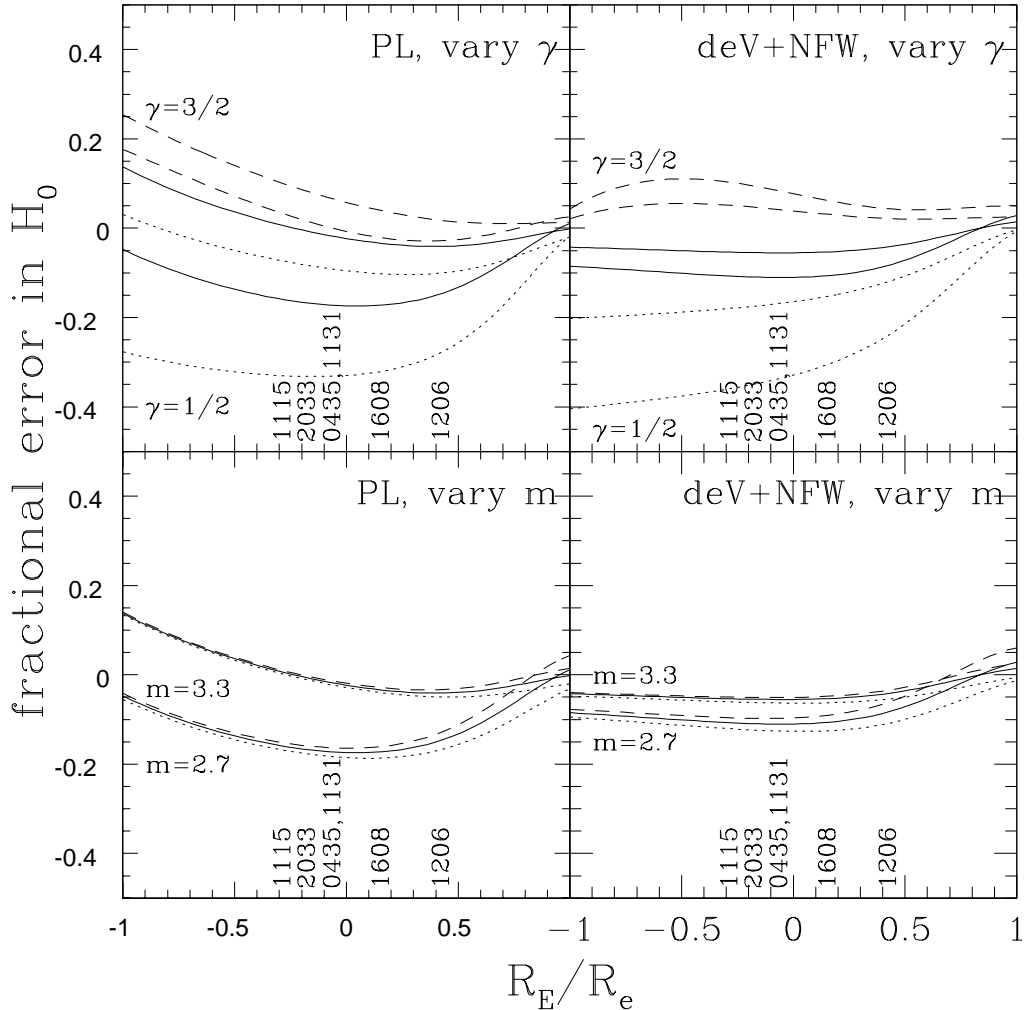}
\caption{ Fractional errors in $H_0$ for an input deV$+$gNFW halo
  modeled as a power law (``PL'', left panels) or a deV$+$NFW halo model
  (deV$+$NFW, right panels).  In the top panels, the asymptotic
  slope $m=3$ is fixed and the inner density slope 
  exponent of the gNFW model is $\gamma=1/2$ (dotted), $\gamma=1$ (solid) 
  or $\gamma=3/2$ (dashed).  In the lower panels, the inner exponent 
  is fixed to $\gamma=1$ while the asymptotic slope is
  $m=2.7$ (dotted), $m=3$ (solid) and $m=3.3$ (dashed).
  The scale length is fixed to $a/R_e=10$ for
  both the gNFW and NFW models and the dark matter fraction is
  either $f_{DM}=25\%$ or $50\%$, with larger fractional errors
  for larger $f_{DM}$.  
  }
\label{fig:gnfw1}
\end{figure*}

\begin{figure*}
\centering
\includegraphics[width=0.80\textwidth]{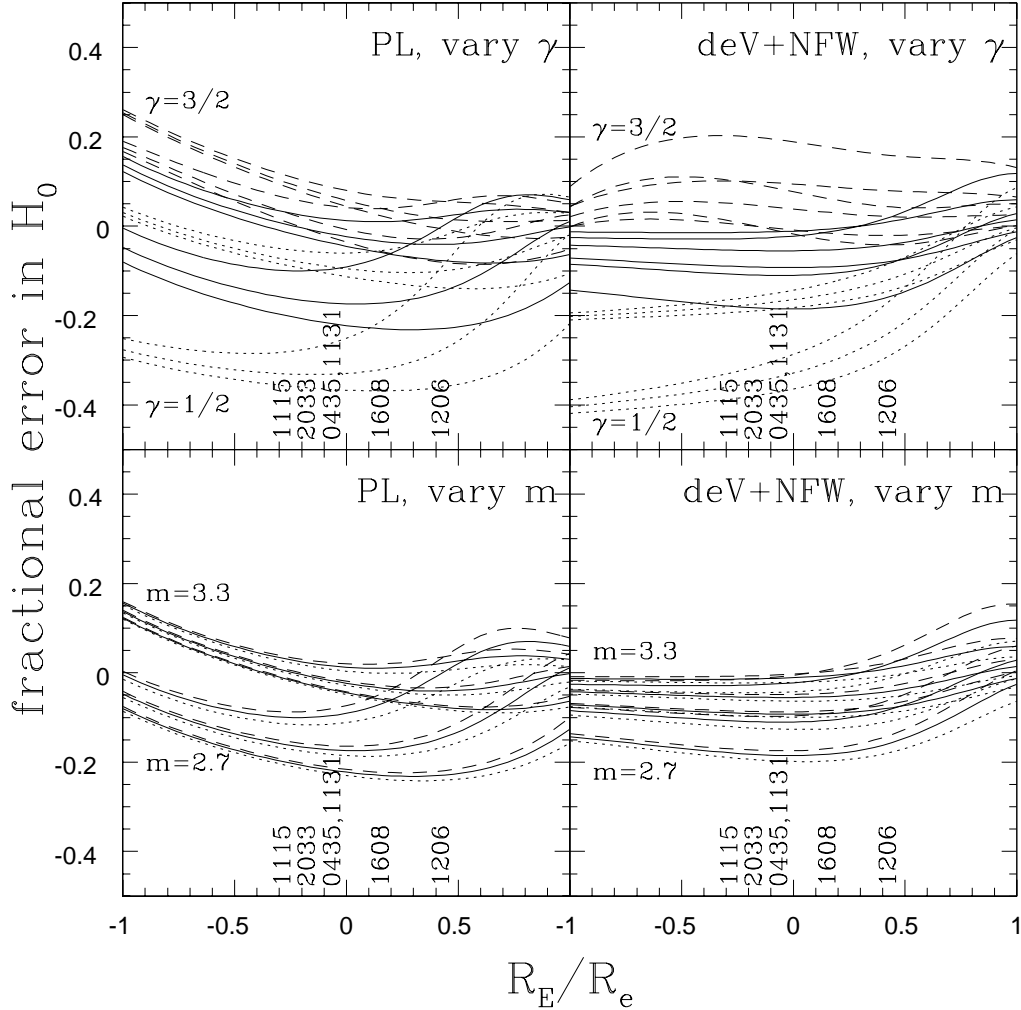}
\caption{ As in Fig.~\protect\ref{fig:gnfw1} but adding the results
  for both models having $a/R_e=5$ and $20$.  The same halo scale length
  is still used for both input gNFW and model NFW profiles.  The 
  spread increases further if we allow the scale lengths to differ
  as in Fig.~\protect\ref{fig:nfw}.
  }
\label{fig:gnfw2}
\end{figure*}

\begin{figure*}
\centering
\includegraphics[width=0.80\textwidth]{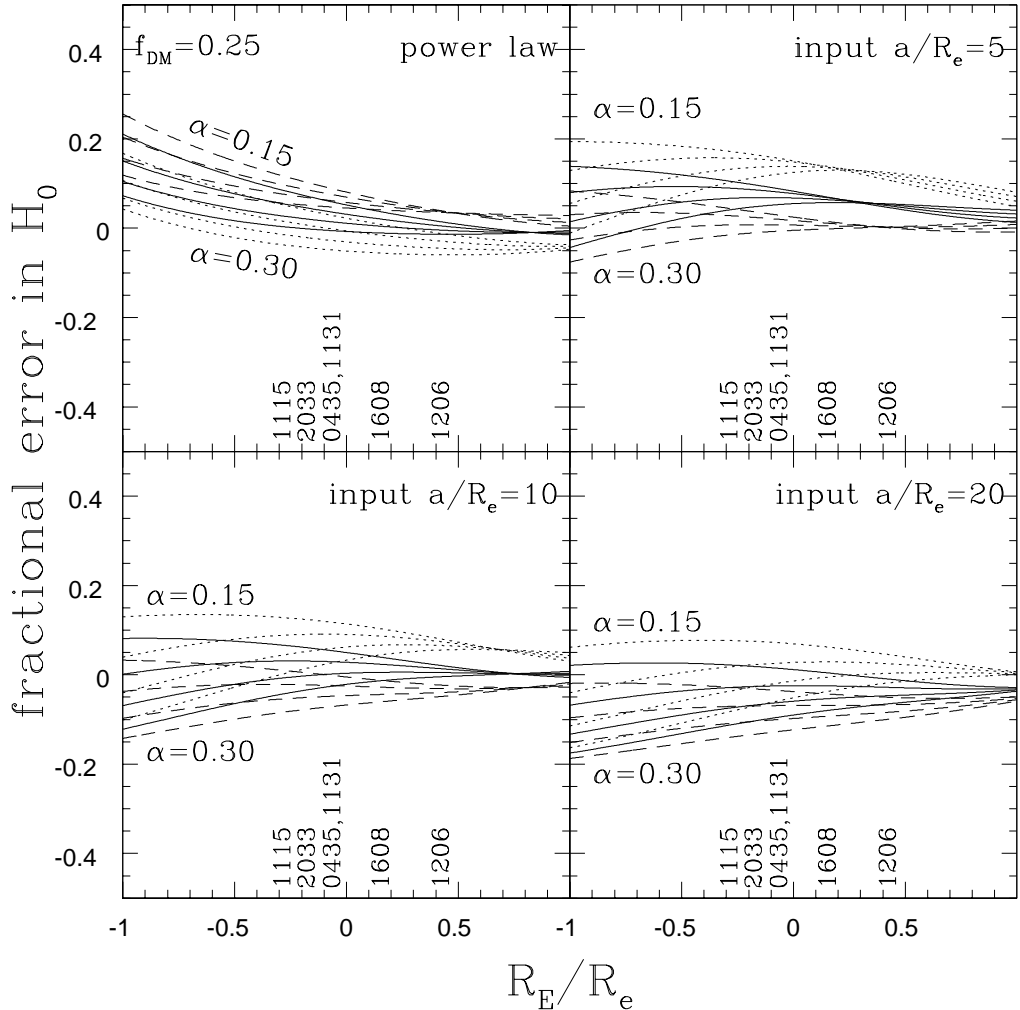}
\caption{ Fractional errors in $H_0$ for deV$+$Einasto models with 
  $f_{DM}=25\%$ and $\alpha=0.15$, $0.20$, $0.25$ and $0.30$. The
  top left panel uses power law models for Einasto halos with 
  $a/R_e=5$ (dashed), $10$ (solid) and $20$ (dotted). The remaining
  panels show deV$+$NFW models applied to Einasto profiles with
  $a/R_E=5$ (top right), $10$ (lower left) and $20$ (lower right).
  The NFW models use $a/R_e=5$ (dashed), $10$ (solid) and
  $20$ (dotted).
  }
\label{fig:einasto}
\end{figure*}

\begin{figure*}
\centering
\includegraphics[width=0.80\textwidth]{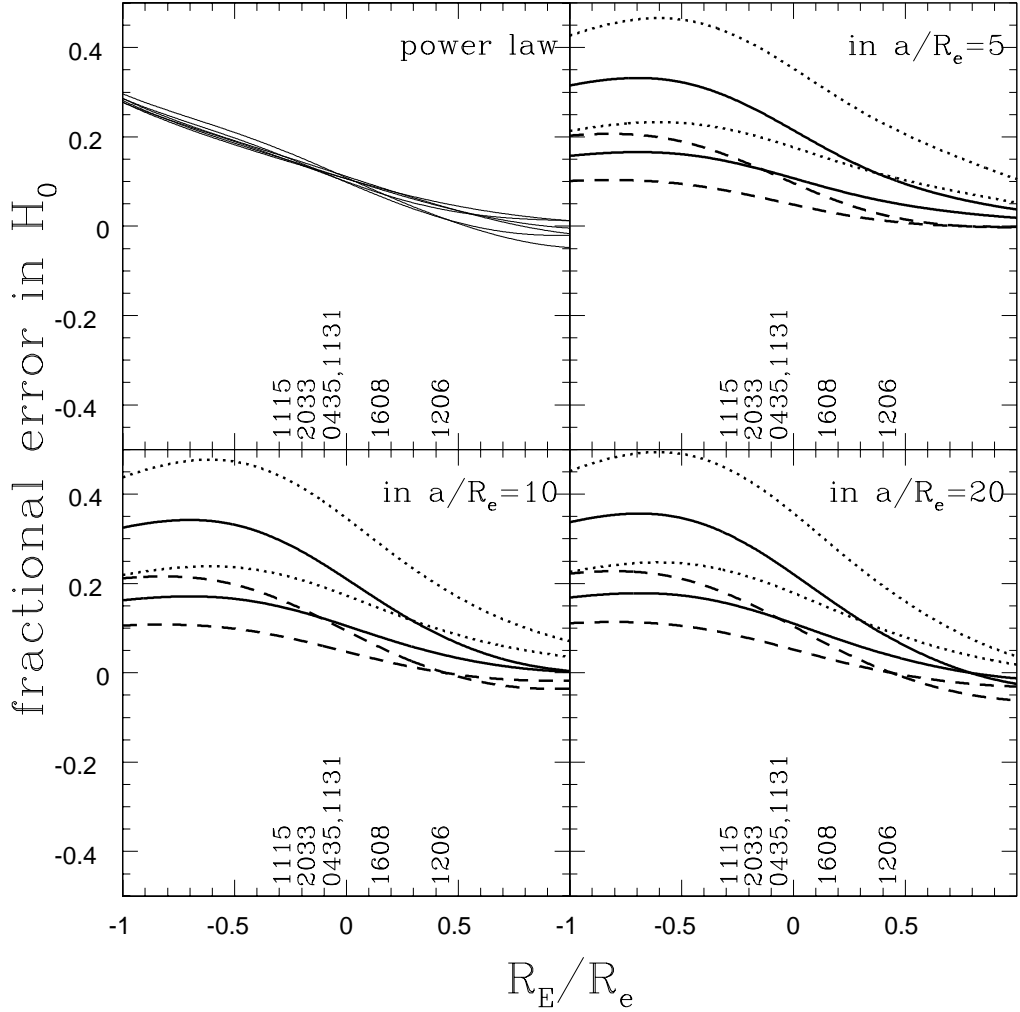}
\caption{ Fractional errors in $H_0$ for adiabatically compressed
  NFW halos.   The top left panel shows the results for the power
  law models and the other three panels are for the 
  deV$+$NFW models where the input model has 
  $a/R_E=5$ (top right), $10$ (lower 
  left) or $20$ (lower right) and the lens model has $a/R_E=5$ (dashed),
  $10$ (solid) or $20$ (dotted). 
  All panels include both the $f_{DM}=25\%$ and $50\%$ cases.
  }
\label{fig:compnfw}
\end{figure*}

\begin{figure*}
\centering
\includegraphics[width=0.80\textwidth]{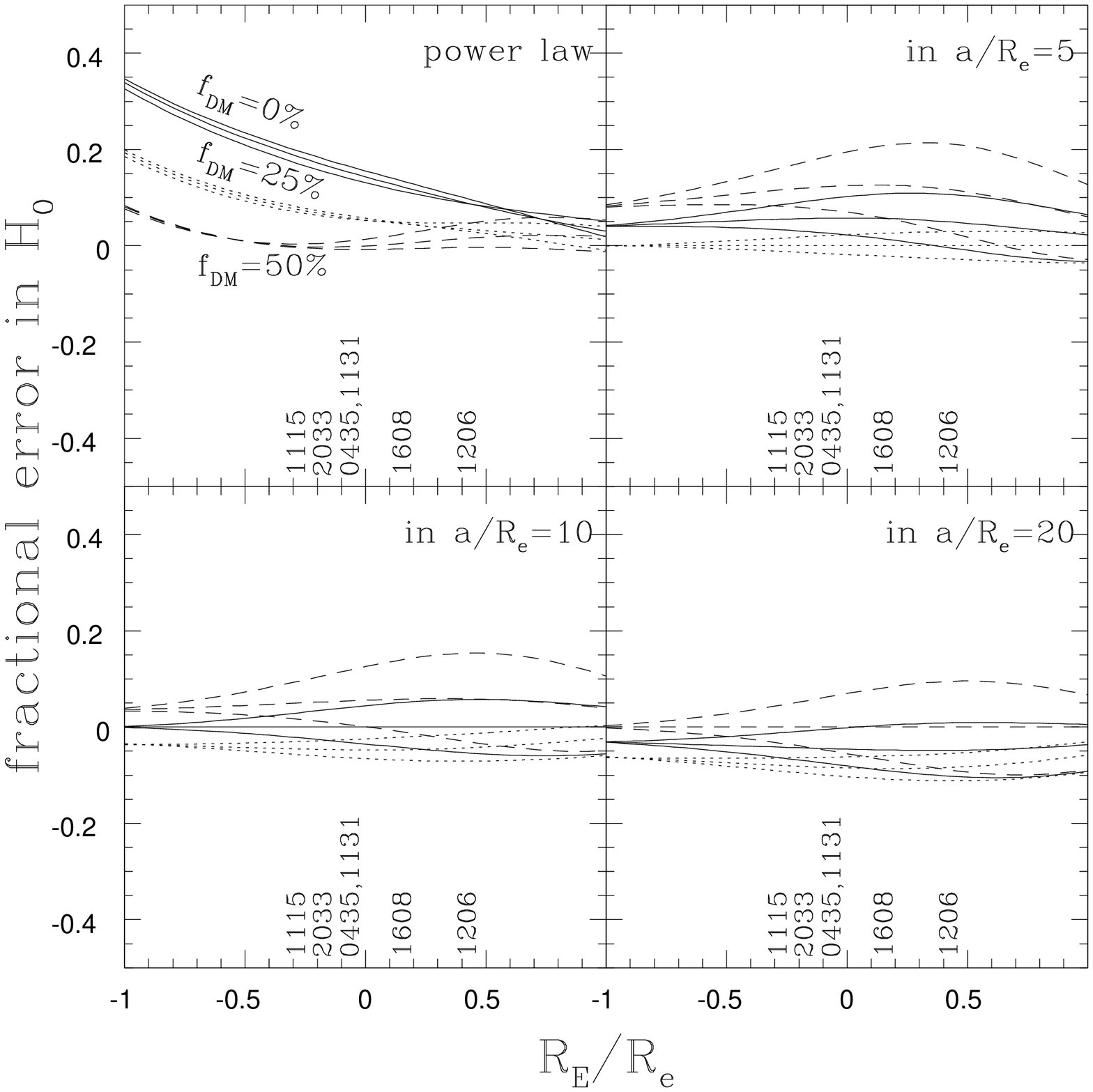}
\caption{ Fractional errors in $H_0$ for models with mass-to-light ratio gradients.
  Results are shown for fractional changes per $R_e$ of $\mu=-0.2$ (top), $0$ and 
  $0.2$ (bottom).  The upper left panel for the power law models shows the cases
  with $f_{DM}=0\%$, $25\%$ and $50\%$.  The other three panels are for the 
  deV$+$NFW models where the input model has $f_{DM}=25\%$ and 
  $a/R_E=5$ (top right), $10$ (lower 
  left) or $20$ (lower right) and the lens model has $a/R_E=5$ (dashed),
  $10$ (solid) or $20$ (dotted).  Generally the $\mu=0.2$ case is at the
  top and the $\mu=-0.2$ case is at the bottom.
  }
\label{fig:ml}
\end{figure*}

\subsection{Consequences}

By matching lens models in $\xi$, it is now easy to show the consequences
of using different mass models in the case of circular lenses.  We used
an input mass distribution consisting of a \cite{deVaucouleurs1948} model for the
stars plus a dark halo.  Qualitatively similar results are obtained if we
chose a different density distribution for the stars. 
We scaled everything by the effective radius $R_e$ 
of the deV model and generated models with 0\%, 25\% or 50\% of the mass 
inside the Einstein radius $R_E$ coming from the halo.  We considered
four halo models.  The first is simply the NFW model of Eqn.~\ref{eqn:nfwprof}.  
The second is the generalization of the NFW (gNFW) profile
\begin{equation}
   \rho \propto { 1 \over r^\gamma \left( a^2+ r^2\right)^{(m-\gamma)/2} }
   \label{eqn:profile}
\end{equation}
introduced as a lens model by \cite{Munoz2001}. This asymptotically matches 
generalizations of the NFW model at large and small radii, but the change in 
structure near the break radius makes the deflection profiles analytic. 
The case $\gamma=1$, $m=3$ is similar to the NFW model, while $\gamma=3/2$,
$m=3$ is similar to the model favored by \cite{Moore1999}.  The third
is the \cite{Einasto1965} profile,
\begin{equation}
   \rho \propto \exp\left[ - { 2 \over \alpha } \left( \left( { r \over a} \right)^\alpha -1 \right)\right]
\end{equation}
where $0.15 \ltorder \alpha \ltorder 0.30$ models may better fit halo 
simulations than the NFW model (e.g., \citealt{Merritt2005}, \citealt{Navarro2010},
\citealt{Reed2011}).  The \cite{Einasto1965} models are most
easily treated numerically.  

These first three models are for dark matter halos unaffected by baryons, but the actual 
halo structures of galaxies are modified by the presence and evolution of the baryons.
In particular, the baryons adiabatically compress the dark matter orbits as they cool
and shrink relative to the dark matter.  As a fourth halo model, we use the simple 
model of adiabatic compression from \cite{Blumenthal1986}.  
We start with an NFW halo and make the final distribution of the baryons a 
\cite{Hernquist1990} density profile with the 
same effective radius as the deV model we use for the lens model. 
We use a NFW concentration of $c=10$, so that the virial radius is $r_v=c a = 10 a$,
and a baryonic mass fraction of 15.7\% (\citealt{Planck2018}).  We then combine this
adiabatically compressed NFW profile with the deV model for the stars, again assuming
that either 25\% or 50\% of the projected mass inside the Einstein ring comes from
the halo -- we did not force the dark matter fraction implied by the adiabatically
compressed model.  

Following H0LiCOW, we model the input system using either the power-law mass 
distribution (``PL'') or the input stellar distribution, here a \cite{deVaucouleurs1948} model,
combined with an NFW model for the halo (``deV$+$NFW'').  We assumed that the effective
radii of the two deV models were fixed and identical.    
With the break radius $a$ of the NFW model fixed, both mass models have
two parameters which we determine by matching the
Einstein radius and $\xi$ of the input model as function
of the Einstein radius relative to the effective radius $R_E/R_e$.  Given
the input $\kappa_{input}$ and model $\kappa_{model}$ surface densities
at the Einstein radius, we can then compute the fractional error in $H_0$ as 
\begin{equation}
    f = {H_{true} \over H_{model}} -1 
   = { 1-\kappa_{input} \over 1 - \kappa_{model}} - 1.
\end{equation}
This has the sense that the models underestimate (overestimate) $H_0$ if 
$f>0$ ($f<0$). 

We first consider models where the input halo is NFW using input break 
radii of $a/R_E=5$, $10$ and $20$.   
H0LiCOW sets $a \simeq (58 \pm 8)h^{-1}$~kpc 
based on the stacked weak lensing analysis of the Sloan ACS lens
sample by \cite{Gavazzi2007}. This roughly
roughly corresponds to $a/R_e \simeq 10$ for most of the H0LiCOW lenses.
Whether from \cite{Gavazzi2007} or simulations (e.g., \cite{Bullock2001}
\citealt{Reed2011}, \citealt{Dutton2014}),
$a/R_e \simeq 10$ is roughly the correct scale.  However, while the 15\%
uncertainty in $a$ found by \cite{Gavazzi2007} and used by H0LiCOW
may be a realistic estimate of the uncertainty in the mean scale
length, it greatly underestimates the plausible range of
scale lengths for individual lenses. The lens galaxies have a finite
spread in halo mass,  and halo concentrations have significant scatter at 
fixed halo mass (e.g., \citealt{Dutton2014}). There are further dependencies on
the redshifts of formation and observation.  Hence, the factor of
two range around $a/R_e=10$ we use for illustration is relatively
realistic even if $a/R_e=10$ is the true mean halo scale length of lenses.
Figure~\ref{fig:nfw} shows fractional errors from modeling these
lenses using either a power law model or deV$+$NFW models with the
same three break radii.  

If lenses happen to have deV$+$NFW mass distributions with $f_{DM}=25\%$
and $a/R_E=10$, then the power law models do remarkably well, with 
fractional errors of only 1-2\% for the range of $R_E/R_e$ spanned
by the lenses.  However, for any other dark matter fraction or 
scale length, the fractional errors quickly exceed 5\%.  The exact
values of the systematic errors found for the power law models are 
quite sensitive to changing the stellar mass distribution. For example,
for this deV$+$NFW model, the fractional error for $R_E=1.3s=0.72R_e$
is 15\% instead of the 30\% fractional error we found for the 
same Einstein radius in the \cite{Hernquist1990}$+$NFW model of
\S2.2.   

The other three
panels of Fig.~\ref{fig:nfw} show the results for the deV$+$NFW model
and the consequences of differences in
the NFW break radius.  If the input and model break radii match,
or there is no dark matter ($f_{DM}=0\%$), then the lens model
can exactly reproduce the input model and the fractional errors
are zero.  However, if the lens model scale length is greater (less)
than the true scale length, $H_0$ is underestimated (overestimated)
with the magnitude of the error increasing with the dark matter
fraction.  Changing the stellar distribution, but still using the
same stellar mass distribution to both generate and model the lens,  
seems to have little effect on systematic errors found when modeling
the system by the stellar density plus an NFW halo.

Next we consider the gNFW models (Eqn.~\ref{eqn:profile}),
where we can vary the inner ($\gamma$) and outer ($m$) logarithmic slopes of
the profile as well as the scale length $a$.  Figure~\ref{fig:gnfw1} shows the results
where both the gNFW and NFW profiles have $a/R_e=10$.
The first point to note is that even with the 
same scale length and exponents matching those of the NFW profile ($\gamma=1$, $m=3$),
there are significant changes in the fractional errors for $H_0$ whether using
the power law or deV$+$NFW models.  As before, the shifts increase as the dark
matter fraction increases.  Varying the outer slope $m$ has relatively little effect
on the results for the $2.7 \leq m \leq 3.3$ range shown.  Varying the inner slope
over the range $1/2 \leq \gamma \leq 3/2$
creates quite large shifts, where the models tend to underestimate (overestimate) 
$H_0$ as we make the inner profile steeper (shallower). The limit $\gamma =3/2$
is the slope favored by \cite{Moore1999}.  As shown in Fig.~\ref{fig:gnfw2},
changing the scale length $a$ to $a/R_e=5$ or $a/R_e=20$ produces significant
changes compared to $a/R_e=10$ even though we continue to use the same break
radius for both the input gNFW mass model and the lens NFW model.  We do
not show the cases where we allow the two break radii to differ, but this leads
to still broader ranges for the fractional errors that are
qualitatively similar to the effects for the deV$+$NFW models
in Fig.~\ref{fig:nfw}.    

Fig.~\ref{fig:einasto} shows the results for the \cite{Einasto1965} halo models
with a dark matter fraction of $f_{DM}=25\%$.  The fractional errors depend on
the parameter $\alpha$, shifting towards more positive fractional errors
as $\alpha$ is 
reduced.  As with the other halo models, more compact halos and halos of one
scale length modeled by one with a smaller scale length are also shifted towards
more positive fractional errors.  The typical scale of the systematic errors
for $f_{DM}=25\%$ is again of order 10\% for reasonable ranges of the model
parameters, rising to $\sim20 \%$ for $f_{DM}=50\%$.

Fig.~\ref{fig:compnfw} shows the results for the adiabatically
compressed NFW halos with a dark matter fraction of $f_{DM}=25\%$. The adiabatically
compressed halos are more centrally concentrated, so it is not surprising that
the main qualitative change from the NFW models in Fig.~\ref{fig:nfw} is to shift 
the fractional errors to larger positive values.  The qualitative shifts 
seen in Fig.~\ref{fig:compnfw} are also found if we adiabatically compress
the \cite{Einasto1965} profiles and are presumably generic.  

So far, we have assumed that the shape of the stellar density distribution
is exactly the same in both generating and modeling the lens, leaving only
the mass-to-light ratio as a parameter of the lens models. Photometric models
of the lens galaxies generally leave small fractional residuals, so if the
stellar distributions have constant mass-to-light ratios this is likely a
safe assumption until pursuing $\sim 1\%$ fractional uncertainties in $H_0$.
However, it is routine to find that surface brightness profiles depend on
the filter of observation or equivalently that early-type galaxies have color gradients 
indicative of radial changes in age or metallicity that in turn imply
changes the stellar mass-to-light ratio (see, e.g., the review by \citealt{Kormendy1989})
Thus, as a final experiment, we gave the input stellar mass distribution a gradient in 
its mass-to-light ratio.  We multiplied the input deV density distribution by
\begin{equation}
     1 + \mu \left( { R - R_e \over R_e }\right)
\end{equation}
but modeled the stellar mass distribution using just the input deV density 
distribution (i.e. $\mu\equiv0$).
For illustration we used $\mu = \pm 0.2$, so a 20\% change in the mass-to-light
ratio per effective radius. We did not 
worry about the mass-to-light ratio becoming negative for large radii when
$\mu < 0$, as all that matters is the mass-to-light ratio from the center to $R_E$,
and the Einstein radii are well inside the radius where the model becomes 
problematic.  As shown in Fig.~\ref{fig:ml}, modest gradients in the
stellar mass-to-light ratio can easily lead to 5-10\% systematic errors in 
estimates of $H_0$ even if the photometric profile of the lens in some
filter is exactly known.  

\section{Discussion}
\label{sec:conclude}

Estimates of $H_0$ from lens time delays are controlled by the convergence 
(surface density)
$\kappa_E$ at the Einstein radius $R_E$, with $H_0 \propto 1-\kappa_E$. No
differential lens data (image separations, flux ratios, etc.) other than
the time delays {\it ever}
directly constrains $\kappa_E$ -- it is a fundamental degeneracy in the mathematics
of lensing (see, e.g., \citealt{Gorenstein1988},
\citealt{Kochanek2002}, \citealt{Kochanek2006}, \citealt{Schneider2013},
\citealt{Wertz2018}).  Lens data constrains two properties of the radial mass distribution:
(1) the Einstein radius $R_E$; and (2) the dimensionless number  
$\xi = R_E \alpha''(R_E)/(1-\kappa_E)$ where $\alpha''(R_E)$ is the second derivative 
of the deflection profile at the Einstein radius.  Any lens with constraints
from more than one set of lensed images will strongly constrain $R_E$ and
$\xi$.  If the (radial) mass model has only two parameters, this will also
lead to tight constraints on $\kappa_E$ and hence $H_0$ because the model
has no additional degrees of freedom.  For example, in power-law lens models
with deflection profiles $\alpha(R) = b^{n-1} R^{2-n}$, $R_E= b$, 
$\xi=2(n-2)$ and $\kappa_E=(3-n)/2=(2-\xi)/4$. But the constraint on 
$\kappa_E$ is purely dictated by the mathematical structure of the
lens model and not by the lens data.
We demonstrate this point in detail for a particular model, admittedly
chosen to lead to an alarming, 30\% fractional error in $H_0$.

We carried out an extensive survey of the consequences of using strong
lens constraints by simply matching $R_E$ and $\xi$ between mass models.
For the input models, we considered lenses consisting of a \cite{deVaucouleurs1948} 
model combined with a broad range of physically reasonable halo models
(the \cite{Navarro1997} NFW model, generalizations of the NFW model,
the \cite{Einasto1965} model, and an adiabatically compressed NFW model).
We then determined the corresponding best fit that would be found using 
the two standard H0LiCOW lens models: the power
law model or the combination of the input \cite{deVaucouleurs1948} model
with an NFW halo.  From the difference between the true and model values
of $\kappa_E$ we can estimate the resulting fractional error in $H_0$.
The typical scale of the systematic error in $H_0$ is $\sim 10\%$.  On
the one hand, this seems remarkably good given the simplicity of the 
mass models.  On the other hand, it
also means that the accuracy of all current estimates of $H_0$ from 
gravitational lens time delays is $\sim 10\%$ independent of the reported
precision of the measurement.  

As emphasized by \cite{Schneider2013}, using mass models with additional
degrees of freedom, so that determining $R_E$ and $\xi$ does not force
a particular value of $\kappa_E$ in our language, is the easiest way to ensure that the
precision of the measurement does not exceed the accuracy even in the
presence of very strong constraints from the lens data.  The power law
model should clearly simply be abandoned -- while it is adequate for
$\sim 10\%$ estimates of $H_0$ it is essentially useless if higher 
accuracies are needed.  Combining the stellar distribution with an 
NFW model can capture much of this uncertainty if the scale radius of
the NFW component is allowed a significant dynamic range.  The current
H0LiCOW models generally constrain the scale length to $10$-$15\%$, 
essentially making it a two parameter model like the power law models.
Even to the extent that NFW models are correct, the scatter of lenses
in mass and the spread of concentrations seen at fixed mass mean that
the scale length should really be allowed to vary by a factor of $\sim 2$.  

While we have emphasized the radial structure of the density distribution
because it then allows us to carry out a large model survey, one should
have similar concerns about the number of degrees of freedom in the 
angular structure.  In our example from \S2.2 of a lens producing a large fractional
error in $H_0$, the problems only worsened when we considered a non-circular
version of the same lens.  The angular structure of the lens drove the models 
to have a radial density distribution with $\kappa_E$ even more divergent from the 
true value than in the circular models.  Models need to have enough
angular degrees of freedom that the angular structure beyond the
quadrupole is not largely determined by the radial mass distribution
of a single ellipsoidal density distribution (see \citealt{Kochanek2006}).  

There will remain a fundamental problem.  While mass models with more degrees
of freedom can capture the uncertainties in $H_0$ created by the uncertainties
in halo structure, these are largely systematic rather than random problems.
For example, if halos were truly NFW models with a factor of two random
scatter in the NFW scale length, then we might legitimately average the
results from multiple lenses to produce a joint estimate of $H_0$ with 
smaller uncertainties than for the individual lenses.  However, if the 
freedom from allowing a broad range of scale lengths is really compensating
for the fact that the real mass distribution is systematically different
from the mean behavior of the model, then there is no reduction in the 
uncertainties from averaging multiple lenses.  

\section*{Acknowledgments}

The author thanks the H0LiCOW collaboration for answering many questions,
and C. Keeton for rapidly fixing a minor problem in {\tt lensmodel}.
CSK is supported by NSF grants AST-1908952 and AST-1814440.

\end{document}